\documentclass[twocolumn,showpacs,showkeys,preprintnumbers,amsmath,amssymb,superscriptaddress]{revtex4}
\usepackage[dvips]{graphicx}
\usepackage{dcolumn}
\usepackage{bm}
\usepackage[usenames]{color}
\usepackage{verbatim} 
\newcommand{\affA}{%
    Van der Waals-Zeeman Institute, University of Amsterdam,
Valckenierstraat 65-67, 1018 XE Amsterdam, The Netherlands }

\newcommand{\affB}{%
   QUANTOP - Danish Quantum Optics Center, Niels Bohr
Institute, 2100 Copenhagen, Denmark}

\def\beq{\begin{equation}}
\def\eeq{\end{equation}}

\begin{document}

\preprint{}

\title{Three-dimensional character of atom-chip-based rf-dressed potentials}

\author{J.~J.~P. van Es}\affiliation{\affA}
\author{S. Whitlock}\affiliation{\affA}
\author{T. Fernholz}\affiliation{\affA}\affiliation{\affB}
\author{A.~H. van Amerongen}\affiliation{\affA}
\author{N.~J. van Druten}\affiliation{\affA}

\date{February 4, 2008} 

\begin{abstract}
We experimentally investigate the properties of radio-frequency-dressed
potentials for Bose-Einstein condensates on atom chips. The
three-dimensional potential forms a connected pair of parallel
waveguides. We show that rf-dressed potentials are robust against the
effect of small magnetic-field variations on the trap potential.
Long-lived dipole oscillations of condensates induced in the rf-dressed
potentials can be tuned to a remarkably low damping rate. We study a
beam-splitter for Bose-Einstein condensates and show that a propagating
condensate can be dynamically split in two vertically separated parts
and guided along two paths. The effect of gravity on the potential can
be tuned and compensated for using a rf-field gradient.
\end{abstract}



\maketitle

\section{Introduction}\label{sec:introduction}

Microscopic magnetic traps for neutral atoms and Bose-Einstein
condensates (BECs) offer a degree of design flexibility that is
unmatched by conventional trapping techniques
\cite{FolKruHen02,Rei02,ForZim07}. An `atom chip' typically
consists of a lithographically defined wire or magnetization
pattern on a substrate to create stable and tailor-made magnetic
potentials for controlling atomic motion on the micrometer scale.
Already important new applications for atom chips have been
demonstrated, including integrated atomic
clocks~\cite{TreHomRei04}, atom
interferometers~\cite{WanAndWu05,ShiSanPre05,SchHofKru05,JoChoPri07}
and BEC-based precision
sensors~\cite{JonValHin03,LinTepVul04,GunKemFor05,WilHofSch05,WilHofBar06,HalWhiSid07},
and soon entire networks of atom-optical elements such as atom
lasers, single-mode atomic waveguides and atomic beam splitters
may be integrated on a single device. The success of atom chips is
due primarily to the close proximity of the atoms to the
field-producing elements ($\lesssim100~\mu$m), which allows high
field gradients and small-scale potential landscapes. As a result,
atom chips are also versatile tools for fundamental studies of
quantum gases in low-dimensions~\cite{HofLesSch07,AmeEsWic07} and
other exotic potentials.

The flexibility of atom chips is increased further by introducing
radio-frequency (rf) oscillating fields to produce rf-dressed adiabatic
potentials for
atoms~\cite{ZobGar01,KetDru96a,ColKnyPerr04,WhiGaoDem06,SchHofKru05,HofLesSch06,LesHofSch06}.
This has recently allowed for the coherent splitting of BECs in
time-dependent double-well potentials~\cite{SchHofKru05,JoShiPre07},
species-dependent microtraps~\cite{ExtLeBSch06}, interferometric
studies of phase fluctuations in the one-dimensional
regime~\cite{HofLesSch07}, and the direct observation of effects beyond
the rotating wave approximation~\cite{HofFisLes07}.

\begin{figure}
\includegraphics[width=0.8\columnwidth]{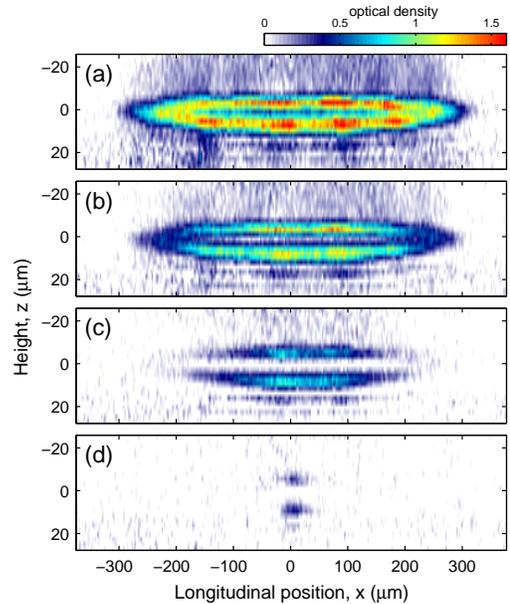}
\caption{(Color online) Resonant absorption images of ultracold $^{87}$Rb atom clouds trapped in a rf-dressed potential. The final temperature of the cloud is varied by varying the final trap depth during evaporative cooling before transferring the atoms to the rf-dressed potential. The final trap depths are (a) 48~$\mu$K, (b) 29~$\mu$K, (c) 9.6~$\mu$K and (d) 2.4~$\mu$K. The observed elliptical distribution of atoms results from the three-dimensional character of the potential. Near the center, $x=0$, the dressed potential forms two parallel waveguides which merge toward the edges of the image at $|x|\approx250~\mu$m.}%
\label{fig:ringpotential}
\end{figure}

In this paper we investigate the properties of atom-chip-based
rf-dressed potentials for Bose-Einstein condensates. Previous studies
of the \emph{transverse} potential have been made, for example, in the
context of double-well BEC
interferometry~\cite{SchHofKru05,HofLesSch06,JoShiPre07}, while a
recent study on the relative phase evolution of two dressed Bose gases
in the one-dimensional regime highlights the importance of the
\emph{longitudinal} degree of freedom in atom chip experiments
\cite{HofLesFis07a}. In general such a rf-dressed potential forms a
three-dimensional structure that can be described as a pair of
connected waveguides as illustrated experimentally in
Fig.~\ref{fig:ringpotential}. Using dressed-state rf
spectroscopy~\cite{GarPerLor06,HofFisLes07} of Bose-Einstein
condensates we characterize and quantify the effect of small magnetic
field variations on the rf-dressed potentials, an effect also recently
studied for chip-based \emph{time-averaged} magnetic
potentials~\cite{TreGarCor07,BouTreGar07}.  We demonstrate a
significant reduction in the effect of magnetic field noise on
rf-dressed potentials and measure an associated reduction of the
longitudinal confinement strength indicating that fluctuating
homogeneous and inhomogeneous fields are effectively suppressed. The
degree of suppression is tuned over one order of magnitude by varying
the dressing radio frequency, providing a new handle for tuning the
strength of disorder potentials on atom chips. We present experimental
results of long-lived longitudinal dipole oscillations of BECs induced
in the rf-dressed potentials and find that the damping can be tuned to
a remarkably low rate. Further, we realize two y-beam-splitters in a
closed loop configuration using the rf-dressed potential. A propagating
BEC can be dynamically split in two vertically separated parts and
guided along two paths. Finally, we show that the effect of gravity on
the symmetry of the potential can be tuned via a rf-field gradient.

This paper is structured as follows. In Sec.~\ref{sec:theory} we
provide the equations describing our rf-dressed potential and present a
model which accounts for a reduction in the potential roughness. In
section~\ref{sec:setup} we describe our experimental setup and the
procedures for preparing Bose-Einstein condensates in rf-dressed
potentials. In section~\ref{sec:results} we present and discuss our
experimental results. Our findings are summarized in
Sec.~\ref{sec:summary}.

\section{Radio frequency dressed potentials}\label{sec:theory}

\begin{figure}[htbp]
\includegraphics[angle=0,width=0.9\columnwidth]{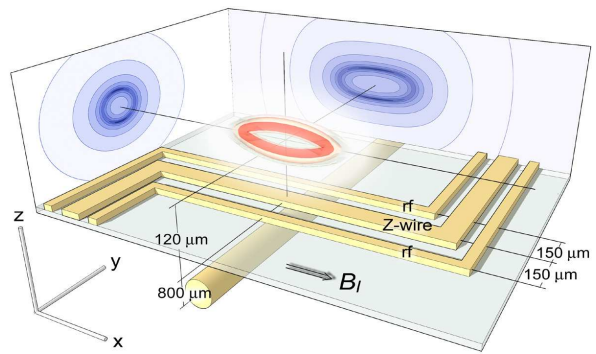}
\caption{(Color online) Schematic of the atom chip used to produce
rf-dressed potentials. The central Z-shaped wire carries a dc-current
and is used together with an external bias field along $y$ to produce
an Ioffe-Pritchard magnetic microtrap. Current through a macroscopic
copper wire beneath the chip surface produces a dip in the longitudinal
magnetic field (along $x$) for additional confinement. Positioned next
to the Z-wire are two wires which carry rf currents. The relative
amplitude of the rf currents are controlled to tune the linear
polarization angle in the transverse $y$-$z$ plane, producing either
horizontally split (along $y$) or vertically-split (along $z$)
rf-dressed potentials. Potential energy cross-sections for vertical
splitting are depicted on the back-planes of the image. A sketch of the trapped atom cloud is shown in red.}%
\label{fig:chip}
\end{figure}

We consider a standard magnetic microtrap (see Fig.~\ref{fig:chip}),
i.e. the field produced by current through a long wire in combination
with an external bias magnetic field which produce a transverse 2D
quadrupole field configuration. Additional wires perpendicular to the
first produce a spatially varying field component oriented along the
wire to provide approximately harmonic longitudinal confinement. The
field components near the magnetic minimum can be conveniently
expressed as:
\begin{equation}\label{eq:field}
\vec{B}=\left(\begin{array}{l}
B_x\\
B_y\\
B_z\\\end{array}\right) \approx\left(\begin{array}{l}
B_{I}+c (x^2-z^2)\\
-q z\\
-q y - 2 c x z\\
\end{array}\right),
\end{equation}
where $B_I>0$ is the Ioffe field offset, $c>0$ is the longitudinal field
curvature, and $q$ is the transverse field gradient.

The coupling of a radio-frequency field to magnetically trapped
atoms depends on the absolute magnetic-field strength, the rf
polarization and on the relative orientation of the static and rf
fields. Following \cite{LesSchHof06}, in the rotating wave
approximation the potential energy experienced by an atom, $U$,
has the simple analytical form:
\begin{eqnarray}\label{eq:U}
U=\tilde{m}\sqrt{\left(|g_F\mu_B\vec{B}|-\hbar\omega\right)^2+\hbar^2\Omega^2},
\end{eqnarray}
where $\tilde{m}$ is the dressed-state quantum number of the atom, $g_F$
is the Land\'{e} g-factor, $\mu_B$ is the Bohr magneton, $\omega>0$ the
frequency of the rf dressing field and $\Omega$ is the position-dependent
Rabi frequency given by the circularly polarized rf-field component
referenced to the local direction of the static magnetic field. In this
section we restrict our analysis to spatially homogenous and linearly
polarized radio-frequency fields with an orientation in the transverse
$y$-$z$ plane. The Rabi frequency is then written as:
\begin{eqnarray}\label{eq:beta}
\Omega
&=&\frac{|g_F\mu_B|}{\hbar}\frac{|\vec{b}_\textrm{rf}\times\vec{B}|}{2|\vec{B}|}\notag\\
&=&\frac{|g_F\mu_B|}{\hbar}\frac{b_\textrm{rf}\sqrt{B_x^2\!+\!\left(B_y\sin\theta\!-\!B_z\cos\theta\right)^2}}{2|\vec{B}|},
\end{eqnarray}
where $\vec{b}_\textrm{rf}$ is the radio frequency field with
amplitude $b_\textrm{rf}>0$ and $\theta$ is the polarization angle
of the radio-frequency field with respect to the $y$ axis. In our
experiments we focus on two configurations, namely $\theta=0$ and
$\theta=\pi/2$. For small dressing frequencies the potential is
approximately harmonic with a single minimum at $x,y,z=0$. At a
critical radio frequency $\omega_0$ the potential splits into two
transversally separated wells. The critical frequency is given by:
\begin{eqnarray}\label{eq:critical-frequency}
\omega_0=\frac{|g_F\mu_B|}{\hbar}\left(B_I -
\frac{b_\textrm{rf}^2}{4 B_I}\right).
\end{eqnarray}
An important quantity is the static magnetic-field strength $B_0$ at
the minimum of the dressed potential. For $\omega\leq\omega_0$ one has
$B_0=B_I$ while for $\omega>\omega_0$, $B_0$ is the positive solution
to the following quartic equation:
\begin{eqnarray}\label{eq:B}
B_0-\frac{\hbar \omega}{|g_F\mu_B|}-\frac{b_\textrm{rf}^2 B_I^2}{4
B_0^3}=0.
\end{eqnarray}
Note that $B_0>B_I$ in the latter case, and $B_0$ depends on the
dressing radio frequency, the rf amplitude and the offset field. The
transverse positions of the minima are found at $y=\nobreak0,z=\nobreak
\pm r_0$ for $\theta=\nobreak 0$ and at $y=\nobreak \pm
r_0,z=\nobreak0$ for $\theta=\nobreak\pi/2$:
\begin{eqnarray}\label{eq:r0}
r_0=\frac{1}{q}\sqrt{B_0^2-B_I^2}.
\end{eqnarray}

Specifically, we are interested in the effect of small spatial
variations of the longitudinal magnetic field on the smoothness of the
rf-dressed potential. This is parameterized as the change in dressed
potential energy as a function of a small change in the magnetic field
$B_I$:
\begin{equation}\label{eq:suppression1}
\frac{dU}{\mu dB_I}=\frac{B_{I}\biggr{[}B_0-\frac{\hbar\omega}{|g_F
\mu_B|}+\frac{b_\textrm{rf}^2}{4B_0}\left(1-\frac{B_I^2}{B_0^2}\right)\biggr{]}}{B_0\biggr{[}\left(\frac{\hbar
\omega}{|g_F\mu_B|}-B_0\right)^2+\left(\frac{b_\textrm{rf}}{2}\frac{B_{I}}{B_0}\right)^2\biggr{]}^{\frac{1}{2}}},
\end{equation}
where $\mu=|\tilde{m}g_F\mu_B|$ is the magnetic moment of the trapped
atom and $B_0$ is given by Eq.~\eqref{eq:B} for $\omega>\omega_0$,
while $B_0=B_I$ for $\omega<\omega_0$. The quantity $dU/\mu d B_I$
(plotted as solid lines in Fig.~\ref{fig:suppression}) is a ratio which
relates the change in potential energy in the rf-dressed potential at
the minimum to a change in the magnetic field strength $B_I$. This
quantity approaches unity for $\omega\ll\omega_0$ where the dressed
potential resembles that of the bare magnetic trap, and decreases
significantly above the splitting point. In the limit of large dressing
frequency $\omega\gg\omega_0$ Eq.~\eqref{eq:suppression1} reduces to:
\begin{equation}\label{eq:suppression2}
\frac{dU}{\mu dB_I}\approx \frac{|g_F\mu_B| b_\textrm{rf}}{2 \hbar
\omega}.
\end{equation}
Hence, for large values of $\omega$, $dU/\mu d B_I$ approaches zero and
the potential energies at the trap minima become effectively insensitive
to small variations of the magnetic field.

\section{Experimental setup}\label{sec:setup}

Our potentials are created using a microfabricated atom chip
produced using optical lithography and metal vapor deposition on a
silicon substrate (Fig.~\ref{fig:chip}).  The chip hosts a series
of 1.8-$\mu$m-thick parallel Z-shaped gold wires with widths
ranging between 5 and 125~micrometers. A current of 1.055~A
through the central, 125-$\mu$m wide Z-wire in combination with an
external bias field of 20~G along $y$ produces an elongated
magnetic trap for $^{87}$Rb atoms at a position of $118~\mu$m from
the chip surface. The magnetic potential is highly elongated with
tight transverse confinement
($\omega_{y,z}\approx2\pi\times880~$Hz). Current through a
macroscopic wire (diameter~0.3~mm) positioned 0.8~mm beneath the
chip surface and perpendicular to the Z-wire produces a dip in the
potential to provide additional longitudinal confinement
($\omega_{x}=2\pi\times48.5~$Hz). The field strength at the
potential minimum is $B_I=2.86$~G. For these trapping parameters
the longitudinal magnetic potential is smooth, with a
root-mean-square roughness below our detection sensitivity of
$\sim1$~mG.

To produce rf-dressed potentials we apply two independent phase-locked
rf currents to wires neighboring the central wire on the chip
(Fig.~\ref{fig:chip}). By varying the relative amplitude of the two
currents we precisely control the orientation of the linearly-polarized
rf field in the plane perpendicular to the chip wires.  For our
experiments we create split potentials oriented in both the horizontal
$y$-direction (parallel to the chip) and the vertical $z$-direction
(against gravity). The rf-field strength is tunable and has a typical
amplitude of $\approx1.7~$G ($1.3$~G) for experiments with horizontal
(vertical) splitting. In our experiments the Larmor frequency of the
atoms is larger than the rf-driven Rabi frequency
($|g_F\mu_B|B_0/\hbar\Omega\geq5$) and we neglect the contribution of
counter-rotating wave terms to the dressed
potential~\cite{HofFisLes07}. The rf-field gradient is significant in
the vertical direction ($\approx18~$G/cm) for vertical splitting and
provides an additional spatial dependence used to tune the asymmetry of
the potential and to compensate the effect of gravity on splitting.

The typical procedure to prepare a Bose-Einstein condensate is as
follows. We start with a cold thermal cloud ($T\approx1~\mu$K) of
$1.5\times10^5$ $^{87}$Rb atoms in the $F=2,m_F=2$ state in a "bare"
static magnetic trap (no rf dressing). To transfer trapped atoms from
the static magnetic trap to the rf-dressed potential ($\tilde{m}=2$) we
typically first switch on the rf below resonance
($\omega/2\pi=1.8$~MHz) and then ramp to the desired dressing frequency
between 1~MHz and 4~MHz. At a critical value of
$\omega_0=1.82$~MHz~($1.90~$MHz) for horizontal (vertical) splitting
the single potential minimum is split in two. Another weak rf field
($\approx15~$mG) is applied to perform forced evaporative cooling in
the dressed potential~\cite{GarPerLor06,HofLesFis06}. The trap depth is
reduced from $4~\mu$K to approximately $1~\mu$K over 350~ms by sweeping
the evaporation frequency from 2.36~MHz to 2.33~MHz. After evaporation
we are left with an almost pure BEC consisting of $3\times10^4$ atoms.
To image the BEC we typically ramp down the rf amplitude to zero within
$20~\mu$s, and then switch off the static magnetic potential to release
the atoms. We use resonant absorption imaging, with the probe oriented
along the $y$-direction (perpendicular to the trap axis, see
Fig.~\ref{fig:chip}).  The typical time-of-flight is 14~ms, the
illumination time is $80~\mu$s and the optical resolution is $4~\mu$m.

\section{Experimental results}\label{sec:results}

Shown in Fig.~\ref{fig:ringpotential} is a sequence of absorption
images taken of atoms trapped in a vertically split rf-dressed
potential for several values for the trap depth. Thermal clouds with
different temperatures are prepared in a bare magnetic potential using
rf evaporative cooling, where the final evaporation frequency
determines the trap depth. The atoms are then transferred to the
rf-dressed potential as described in Sec.~\ref{sec:setup}, and the
dressing radio frequency is linearly ramped from 1.80~MHz to 2.4~MHz in
0.5~s. For final trap depths greater than approximately $25~\mu$K
(corresponding to relatively high-temperature thermal clouds) we
observe a distinctive elliptical distribution of atoms which reflects
the characteristic shape of the dressed potential. This shape and the
observed atomic distribution is due to the inhomogeneous longitudinal
magnetic field which introduces a spatial dependence of the rf
detuning. Near $x=0$ the potential resembles two parallel waveguides
with a vertical separation of approximately $12~\mu$m. For increasing
$|x|$, the local Ioffe field $B_x$ increases and the waveguides
approach each other, joining near $|x|=250~\mu$m where the local Ioffe
field crosses the critical value defined by the radio frequency
[Eq.~\eqref{eq:critical-frequency}]. As the trap depth and the cloud
temperature are reduced further, atom clouds in the two waveguides
disconnect and the atoms remain localized at the two minima of the
potential.

\subsection{Effect of magnetic field variations}
\label{sec:results_smoothness}

\begin{figure}
\includegraphics[width=0.9\columnwidth]{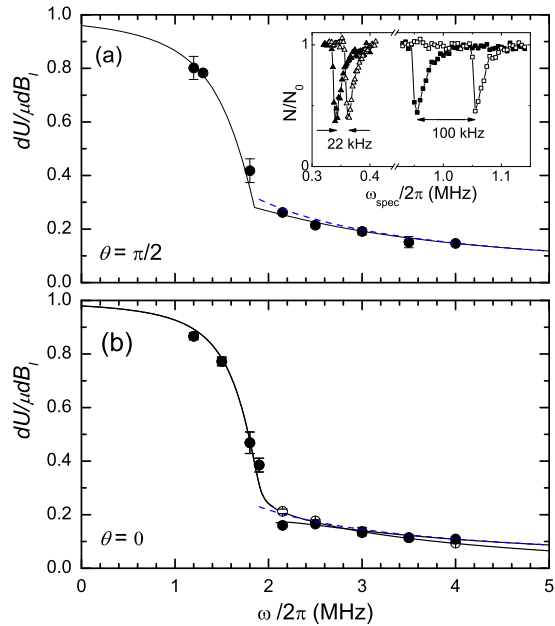}
\caption{(Color online) The suppressed effect of small magnetic
field variations as a function of rf-dressing frequency $\omega$
measured using rf spectroscopy of dressed Bose-Einstein
condensates. Measurements have been performed for both
horizontally split (a) and vertically split (b) potentials. The
inset of (a) shows typical rf spectra for two dressing
frequencies: $\omega=2\pi\times3.5$~MHz (triangles) and
$\omega=2\pi\times1.3$~MHz (squares) for two values of $B_I$
differing by 0.186~G. The measured suppression ratio $dU/\mu dB_I$
is in excellent agreement with the prediction of
Eq.~\eqref{eq:suppression1} (solid lines) and the approximation of
Eq.~\eqref{eq:suppression2} (dashed lines) in the limit of large
dressing frequency. In (b) we also consider the finite rf field
gradient which accounts for the small difference between the two
vertically split traps (solid and open circles).}
\label{fig:suppression}
\end{figure}

First we measure the change in potential energy at the minimum of the
rf-dressed potential due to a small change in the magnetic field
strength $B_I$ oriented along the
$x$-direction~(Fig.~\ref{fig:suppression}). Note that small static
field components oriented in the transverse $y-z$ plane simply displace
the minimum of the static field and their contribution to the potential
energy near the minimum can be neglected. We spectroscopically measure
the change in potential energy by probing the energy difference between
dressed states of a BEC with an additional weak rf field of tuneable
frequency~~\cite{HofFisLes07}. As the weak rf field is tuned in
resonance with the dressed-state level spacing we observe significant
atom loss. Changing $B_I$ affects the potential energy of the trapped
atoms resulting in a measurable shift of the resonant frequency. The
shift is recorded for many values of the rf-dressing frequency and is
then compared to the difference in potential energy associated with a
change in $B_I$ for a bare (undressed) magnetic trap.

The experimental procedure is as follows. A BEC is prepared in a
rf-dressed potential as described in Sec.~\ref{sec:setup}. We linearly
ramp the dressing frequency to a desired value at a relatively slow
rate of 3.3~MHz/s to minimize excitations. Radio-frequency spectroscopy
of the dressed condensate is then performed using a rf pulse, which is
weak compared to the rf dressing field ($\sim0.01 b_\textrm{rf}$). The
duration of the pulse is adjusted depending on the dressing frequency
to remove a large fraction, but not all, of the atoms from the trap at
resonance. Typical rf spectra are shown in the inset of
Fig.~\ref{fig:suppression}(a). Each data point corresponds to a new
experimental run where the spectroscopy frequency
$\omega_{\text{spec}}$ is iterated over a range corresponding to
$|\omega_{\text{spec}}-\Omega|<2\pi\times100$~kHz with a resolution of
a few kHz. The width of the resonance is typically $\approx
2\pi\times10$~kHz and on resonance the spectroscopy field effectively
removes about half the atoms from the BEC. For $\omega>\omega_0$ the
BEC splits in two spatially separated parts. Because of the vertical rf
field gradient, for vertical splitting we measure two resonant
frequencies; one associated with each of the two wells. As a result we
can selectively out-couple atoms from either well. The resonant
frequency provides a measure of the potential energy at the trap
minimum for a given value of $B_I$. To obtain the dependence of the
potential energy on $B_I$ for a particular value of $\omega$, we
measure spectra for five offset field values around $B_I\approx2.86$~G
each separated by 93~mG. Over this small range the resonant frequency
varies approximately linearly with $B_I$ and the slope of a fit to the
measurements gives $dU/\mu dB_I$. This procedure is then repeated many
times for different values of the dressing radio frequency $\omega$.

A full set of measurements for many values of the rf-dressing frequency
and $b_\textrm{rf}=1.7~(1.3)$~G for horizontal (vertical) splitting is
shown in Fig.~\ref{fig:suppression}. For both horizontal and vertical
splitting we measure a reduction in $dU/\mu d B_I$ with increasing rf
dressing frequency. The solid lines show the expected $dU/\mu d B_I$
based on the theory presented in Sec.~\ref{sec:theory}
[Eq.~\eqref{eq:suppression1}] which is in excellent agreement with the
experimental results. The dashed lines indicate the simple
approximation for $\omega\gg\omega_0$ [Eq.~\eqref{eq:suppression2}].
The kink near $\omega/2\pi\approx1.85$~MHz corresponds to the splitting
point ($\omega=\omega_0$), above which the potential has two minima.
Far above the splitting point ($\omega\gg\omega_0$) $dU/\mu d B_I$
decreases as $1/\omega$ [Eq.~\eqref{eq:suppression2}]. Our measurements
at $\omega/2\pi=4$~MHz indicate $dU/\mu d B_I \approx 0.1$,
corresponding to a reduction of the sensitivity to magnetic field
variations by a factor of ten. This factor can be increased further for
even larger dressing frequencies, but is limited in our case by an
associated reduction in the coupling strength (due to the changing
orientation of the local magnetic field at high dressing frequencies)
which leads to non-adiabatic transitions to untrapped dressed states.
We expect that much larger suppression factors ($dU/\mu dB_I\approx0$)
could be obtained for dressed-states with $\tilde{m}<0$ (trapped at the
origin for $\omega>\omega_0$ so that $B_0=B_I$), as
Eq.~\eqref{eq:suppression1} then crosses zero near resonance
($\omega=|g_F\mu_B|B_I/\hbar$). This enhanced magnetic field
suppression may be the topic of future work.

\subsection{Longitudinal dipole oscillations}

In the previous section we have shown that the effect of magnetic
field variations on the rf dressed potential is significantly
reduced. This is of particular consequence in magnetic microtraps
where spatial variations of the longitudinal `Ioffe' field cause
fragmentation of trapped atom
clouds~\cite{LeaChiKie02,OttForKra03,TreGarCor07}. This effect is
suppressed in radio-frequency-dressed potentials which therefore
should allow for exceptionally smooth traps and waveguides. To
investigate this further we observe the dynamical evolution of
BECs propagating in the rf-dressed potential via long-lived
longitudinal dipole oscillations. The spatially inhomogeneous
field which provides longitudinal confinement is reduced for large
rf-dressing frequencies resulting in lower harmonic oscillation
frequencies.

\begin{figure}
\includegraphics[width=0.9\columnwidth]{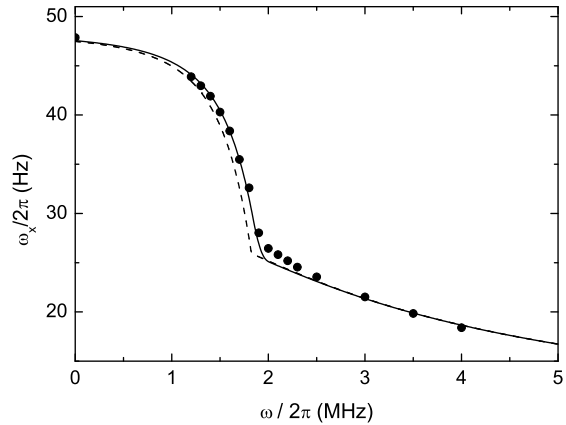}
\caption{Measured longitudinal dipole oscillation frequencies of
Bose-Einstein condensates in radio-frequency-dressed potentials. The
data is obtained for horizontal splitting and $b_\textrm{rf}=1.7$~G.
The dashed line is the calculated decrease in potential curvature at
the trap minimum ($\omega_x'/\omega_x\approx\nobreak\sqrt{dU/\mu d
B_I}$) based on the parameters obtained from the suppression
measurements of Sec.~\ref{sec:results}.A. The solid line includes
corrections due to anharmonicity of the longitudinal potential which
slightly increases the BEC oscillation frequency below resonance.}
\label{fig:axialfreq}
\end{figure}

Longitudinal dipole oscillations of BECs are produced experimentally by
applying a magnetic field gradient along $x$ to displace the potential
minimum. The gradient pulse has a cosine temporal profile with a period
of 16~ms. The gradient is adjusted to produce oscillations with a
roughly constant amplitude between 25 and 50~$\mu$m. After inducing the
oscillation, we apply an additional weak rf knife above the trap bottom
which lowers the trap depth and minimizes heating and the effect of
thermal atoms on the oscillations. In all measurements the BEC remains
pure within our detection sensitivity for its entire lifetime of
approximately 2~s. After various oscillation times we release the BEC
and after 14~ms of ballistic expansion we take an absorption image to
determine the longitudinal center-of-mass position. The BEC position
vs. oscillation time is fit to a exponentially damped cosine function
to determine the oscillation frequency and damping time.

Shown in Fig.~\ref{fig:axialfreq} are the fitted oscillation
frequencies for various values of the rf dressing frequency. The
oscillation frequency decreases as expected from the reduced
confinement strength, $\omega_x'/\omega_x\approx\nobreak\sqrt{dU/\mu d
B_I}$ (Fig.~\ref{fig:axialfreq}-dashed line). Better agreement with the
measured oscillation frequency can be obtained by calculating the mean
curvature~\cite{AntPitStr04} of the rf-dressed potential in the
longitudinal direction weighted by the BEC distribution~
(Fig.~\ref{fig:axialfreq}-solid line), which takes into account the
small longitudinal anharmonicity of the dressed potential in the
vicinity of the splitting point. Our results indicate a significant
reduction in the effect of spatially inhomogeneous magnetic fields in
rf-dressed potentials.

\begin{figure}
\includegraphics[width=0.9\columnwidth]{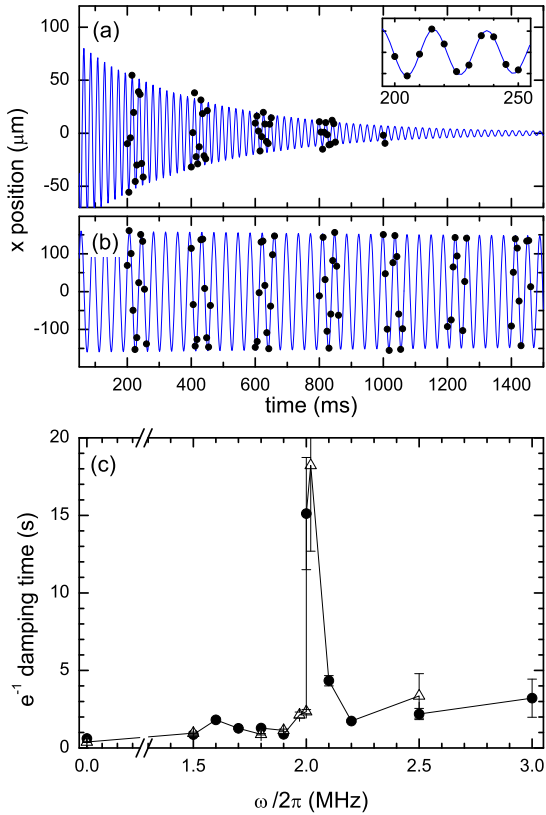}
\caption{(Color online) Damping of longitudinal dipole oscillations in
the rf-dressed potentials.  The center of mass position of a BEC as a
function of time is recorded after 14~ms expansion for (a) the bare
magnetic trap and for (b) a rf-dressed potential with dressing
frequency $2 \pi \times 2.02$~MHz. The inset of (a) shows a zoomed-in
section of the oscillation in the bare magnetic trap. In (c) we show
the measured $e^{-1}$-damping time for various traps as a function of
the rf-dressing frequency. The two datasets (circles/triangles)
correspond to measurements obtained on two different days.}
\label{fig:damping}
\end{figure}

In addition to the oscillation frequency, the damping rates provide a
sensitive measure of the properties of the rf-dressed potential. Shown
in Fig.~\ref{fig:damping} are the $e^{-1}$-damping times obtained from
the above fits to the experimental data as a function of the rf
dressing frequency. In the bare magnetic trap we observe a damping time
of $0.5\pm0.1$~s, significantly less than our observations in the
rf-dressed trap. Below the splitting point we observe a damping time of
$1.2\pm0.4$~s ($\omega<\nobreak\omega_0=\nobreak 2\pi\times1.85$~MHz),
while far above the resonance the damping time is further increased to
$2.4\pm0.8$~s. Remarkably, near resonance
($\omega\approx|g_F\mu_B|B_I/\hbar=\nobreak2\pi\times$2.00~MHz) we
observe effectively undamped oscillations for the entire lifetime of
the BEC consistent with a damping time of $>$10~s, a factor of 20
greater than for the bare magnetic trap. Note that at this point the
potential is split and the barrier height is sufficiently high to keep
the two BECs separated. The exact origin of this sharp decrease in the
damping rate remains unclear. One-dimensional numerical simulations of
condensates oscillating in our potential have been performed by solving
the Gross-Pitaevskii equation for a rough potential (similar to the
calculations described in \cite{TreGarCor07} for a thermal atom cloud),
but are unable to account for the observations given our trap
parameters suggesting a full three-dimensional description is required.
The observed damping may be due to the mixing terms in the potential
(terms in Eq.~\eqref{eq:B} depending on both $x$ and $z$) coupling
longitudinal motion to the transverse degrees of freedom. To first
order these cross-terms can be neglected for $\omega\approx\omega_0$
and this may lead to measurably lower damping rates.

\subsection{Beam-splitters for Bose-Einstein condensates}

Our results highlight the application of atom-chip-based rf-dressed
potentials as smooth atomic waveguides for Bose-Einstein condensates.
These potentials can also act as atomic
beam-splitters~\cite{SchHofKru05,JoShiPre07}, necessary components for
example in an atomic Mach-Zender interferometer. We have studied both
temporal and spatial beam-splitting of stationary and propagating BECs
prepared in our rf-dressed potentials. As the dressing radio frequency
is increased from below to above resonance, the potential is smoothly
transformed from a single well to a double-well potential.
Alternatively, one can make use of the spatial dependence of the rf
detuning due to the inhomogeneous longitudinal magnetic field. In this
case the potential consists of a pair of connected waveguides which
merge at the longitudinal extremes (Fig.~\ref{fig:ringpotential}).

\begin{figure}
\includegraphics[width=0.9\columnwidth]{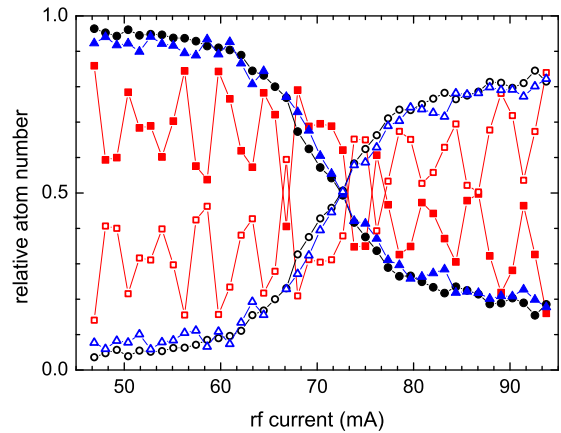}
\caption{(Color online) Effect of potential gradients on the
time-dependent splitting of rf-dressed Bose-Einstein condensates. A BEC
is first prepared in a single well and then the dressing radio
frequency is linearly increased by $0.25$~MHz to split the BEC in the
(vertical) $z-$direction. The rf current is varied to change the rf
field strength and gradient thereby tuning the potential symmetry for
fixed rf polarization. The rf-field gradient in the $z-$direction is
determined by the wire geometry and is calibrated to approximately
0.25~G/mAcm. Solid symbols correspond to the population in the lower
well, open symbols to the upper well. We observe equal splitting for a
rf current of 73~mA ($b'_\textrm{rf}\approx 18$~G/cm) indicating that
the effect of gravity can be compensated. Data is provided for three
rf-ramp durations $t_s$ of 250~ms (triangles), 75~ms (circles), and
2.5~ms (squares).
The complex behavior observed in the 2.5~ms data is most likely
due to collective excitations induced during rapid splitting.}
\label{fig:splittingratio}
\end{figure}

\begin{figure}
\includegraphics[width=0.9\columnwidth]{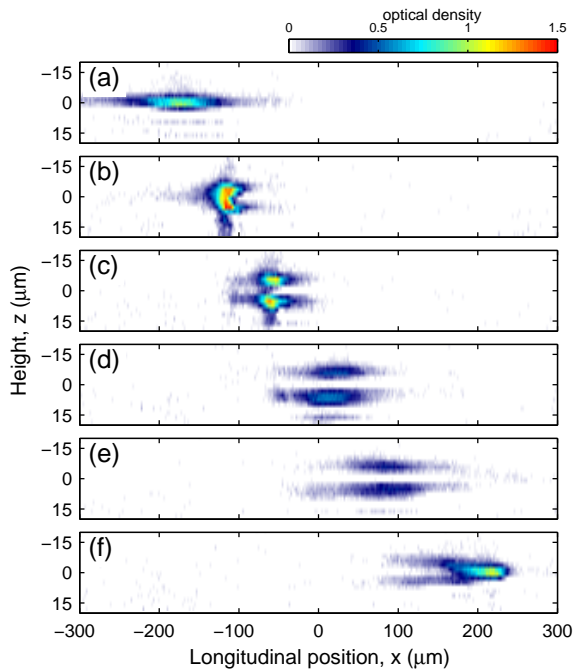}
\caption{A spatial beam-splitter for a propagating Bose-Einstein
condensate. In (a) a single BEC is prepared near the edge of the
potential by exciting a large-amplitude longitudinal oscillation.
At the turning point the parameters of the rf-dressed potential
are switched to create a pair of waveguides. The BEC propagates at
a velocity of 20~mm/s and is split at a position of $x=-160~\mu$m
(b). Two parts of the initial BEC propagate in two vertically
separated waveguides (c)-(e) and recombine after 16~ms at a
position of $x=160~\mu$m (f).} \label{fig:ring-BEC}
\end{figure}

We perform \emph{temporal} splitting of a BEC in an rf-dressed
potential in the vertical $z-$direction by linearly ramping the rf
dressing frequency from 1.80~MHz to 2.05~MHz during a time $t_s$. The
splitting point is between 1.85~MHz and 2.00~MHz depending on the rf
amplitude during the splitting process.  To split a BEC in equal parts
it is necessary to consider perturbations to the rf-dressed potential
due to gravity and field gradients caused by the proximity of the
trapping wires. In our experiments the presence of an Ioffe field
gradient $B'_I=dB_I/dz$ and a rf field gradient
$b'_\textrm{rf}=db_\textrm{rf}/dz$ change the symmetry of the
rf-dressed potential and together act to compensate for the
gravitational asymmetry.  To first order the symmetry of the overall
potential evaluated at the splitting point $\omega_0$ is given by:

\begin{equation}
\frac{dU}{dz}\biggr{|}_{\omega_0}\!\!\!\!=\tilde{m}|g_F
\mu_B|\frac{(b_\textrm{rf} B'_I+B_I
b'_\textrm{rf})}{\sqrt{4B_I^2+b_\textrm{rf}^2}} - M g
\label{eq:gravitycomp}
\end{equation}

where $g$ is the gravitational acceleration and $M$ is the mass of the
Rb atom.  Equation~\eqref{eq:gravitycomp} shows that the field
strengths and gradients $b_\textrm{rf}, B_I, b'_\textrm{rf}$ and $B'_I$
can be used to control the symmetry of the potential at the splitting
point and therefore also play a role in tuning the splitting ratio for
trapped atoms.

In Fig.~\ref{fig:splittingratio} we show the relative number of atoms
in each well after time-dependent splitting a BEC in the rf-dressed
potential. The symmetry of the potential is controlled by changing the
rf currents through the rf wires which affects both $b_\textrm{rf}$ and
$b'_\textrm{rf}$. The static field is kept fixed with a gradient of
$B'_I\approx74$~G/cm. For long splitting times ($t_s\geq75$~ms) the
atoms follow the global minimum of the split potential. For small rf
currents (small $b'_\textrm{rf}$) almost all the atoms of the BEC are
measured in the lower well, while for high rf currents (large
$b'_\textrm{rf}$) gravity is over-compensated and the upper well is
predominately populated. For a critical rf current of 73~mA
corresponding to a rf-field gradient of $b'_\textrm{rf}\approx18$~G/cm
we measure equal populations in each well, indicating the potential is
symmetric during splitting and the effect of gravity can be compensated
effectively. The critical value of $b'_\textrm{rf}$ for equal splitting
is approximately 10~G/cm higher than expected from
Eq.~\eqref{eq:gravitycomp}. We attribute this discrepancy to an
uncertainty in determining the physical splitting point for the BEC
which (due to the finite chemical potential) occurs for dressing
frequencies slightly higher than $\omega_0$, where the effect of $B'_I$
on $dU/dz$ is weaker (similar to the suppression exhibited in
Eq.~\eqref{eq:suppression1}). We have also measured the splitting ratio
for rapid splitting of the potential corresponding to $t_s=2.5$~ms. In
this case we observe less pronounced population imbalances after
splitting and a reduced sensitivity to the symmetry of the potential.
For very short splitting times the final distribution of atoms is
expected to become roughly independent of the symmetry of the
potential~\cite{SidDalSch06}. Additionally we observe complex
oscillatory behavior of the final population imbalance presumably
caused by collective excitations produced during splitting.

To demonstrate a \emph{spatial} beam-splitter for guided atoms we begin
by positioning a Bose-Einstein condensate just outside the split region
of the rf-dressed potential. This is performed in a single potential
well by rapidly reducing the longitudinal magnetic confinement and
displacing the trap minimum by 280~$\mu$m in the $-x$ direction with a
magnetic field gradient. The offset field $B_I$ is also increased. As a
result, the BEC begins a slow longitudinal oscillation with a period of
approximately 120~ms. After 40~ms the field gradient is adjusted to
position the condensate at a turning point of the new potential at a
position of $x=-200~\mu$m. The applied field gradient is then removed
and the longitudinal confinement is increased again within 1~ms.
Simultaneously the offset field is returned to $B_I=2.86$~G. The
dressing rf frequency is $\omega=2\pi\times2.25$~MHz (above resonance)
and creates an elliptically shaped potential, cf.
Fig~\ref{fig:ringpotential}a.

Under the influence of the increased longitudinal confinement the BEC
is accelerated toward the center of the split potential. Absorption
images of the BEC taken {\it in-situ} show the resulting dynamics
(Fig.~\ref{fig:ring-BEC}). After 2~ms of propagation the condensate has
a velocity of 20~mm/s and arrives at the splitting point at $t=0$. We
confirm at this point that the BEC is nearly pure with no thermal
fraction discernable in absorption images taken after time-of-flight
expansion. The shape of the BEC changes dramatically at the splitting
point where it separates in two components, each propagating in
separate and vertically displaced waveguides.  After an additional 2~ms
the propagating clouds appear diffuse in time-of-flight images,
suggesting that the excitations wipe out the BEC. The two separated
components continue to propagate to the end of the potential where they
merge. For longer times the cloud reverses direction, is re-split and
eventually recombines at $x=-200~\mu$m to complete a full oscillation
in the rf-dressed potential. To further improve on this result one
could try to minimize excitations by reducing the velocity of the BEC.
This was not possible in the present work because of the relatively
high gradients in the potential outside the split region. Additionally
the angle under which the waveguides meet in the splitting point is
large, so that the splitting is relatively sudden, probably causing
additional excitations.

\section{Summary and Conclusion}\label{sec:summary}

In summary, we have presented a detailed investigation into many relevant
properties of radio-frequency-dressed potentials on an atom chip.
We have shown that the potential can be described as a pair of
parallel waveguides that connect and merge at the ends,
highlighting the importance of the full three-dimensional
character of these potentials.

We  have demonstrated theoretically and experimentally using
dressed-state radio-frequency spectroscopy, that the effect of
magnetic-field noise can be suppressed by an order of magnitude,
thereby providing a new handle for tuning the strength of disorder
potentials on atom chips. We have studied longitudinal dipole
oscillations of Bose-Einstein condensates in the rf-dressed
potentials. The frequency of these oscillations is reduced, in
agreement with the above suppression. We observe damping times
which dramatically increase from 0.5~s to more than 10~s when the
dressing radio frequency is tuned close to resonance
($\omega=|g_F\mu_B|B_I/\hbar=\nobreak2\pi\times$2.0~MHz). This
unexpected result is not fully understood and deserves more
attention. Long-lived dipole oscillations allow accurate
determination of the trapping potential and are important in novel
applications such as precision measurements of forces near
surfaces \cite{HarObrMcG05}.

We have investigated both temporal and spatial beam-splitters
for Bose-Einstein condensates using the rf-dressed potentials.
We have demonstrated that it is possible to tune the symmetry
of the potentials by varying the rf-field gradient. In this
way the effect of gravity on beam-splitting is compensated.
We have demonstrated that this can alternatively be used to
create  a highly asymmetric beam-splitter. A propagating
Bose-Einstein condensate has been equally split in two parts
and guided along two vertically separated waveguides. The
splitting is accompanied by excitations that quickly destroy
the BEC. We anticipate that this limitation may be overcome
using more sophisticated, custom-designed atom chip potentials.

In conclusion, studies of atom-chip-based rf-dressed potentials as
presented here, highlight new possibilities and opportunities
offered by these potentials. These are relevant to a wide range of
ultracold-atom systems, such as guided-wave atom
lasers~\cite{GueRioGae06}, one-dimensional
gases~\cite{AmeEsWic07,HofLesSch07}, and
interferometers~\cite{HofLesFis07a,WanAndBri05}.

\acknowledgments

We gratefully acknowledge R. J. C. Spreeuw, M. Baranov, I.
Lesanovsky, G. V. Shlyapnikov and J. T. M. Walraven for helpful
discussions. The atom chip was produced using the facilities of
the Amsterdam nanoCenter and with the help of C. R\'{e}tif and J.
R\"{o}vekamp. This work is part of the research program of the
Stichting voor Fundamenteel Onderzoek van de Materie (Foundation
for the Fundamental Research on Matter), and was made possible by
financial support from the Nederlandse Organisatie voor
Wetenschappelijk Onderzoek (Netherlands Organization for the
Advancement of Research) and by the EU under contract
(MRTN-CT-2003-505032).

\bibliography{group-current,swhitlock}

\begin{thebibliography}{10}

\bibitem{FolKruHen02}
R. Folman {\it et~al.}, Adv. At. Mol. Opt. Phys. {\bf 48},  263  (2002).

\bibitem{Rei02}
J. Reichel, Appl. Phys. B {\bf 75},  469  (2002).

\bibitem{ForZim07}
J. Fort\'{a}gh and C. Zimmermann, Rev. Mod. Phys. {\bf 79},  235  (2007).

\bibitem{TreHomRei04}
P. Treutlein {\it et~al.}, Phys. Rev. Lett. {\bf 92},  203005  (2004).

\bibitem{WanAndWu05}
Y.~J. Wang {\it et~al.}, Phys. Rev. Lett. {\bf 94},  090405  (2005).

\bibitem{ShiSanPre05}
Y. Shin {\it et~al.}, Phys. Rev. A {\bf 72},  021604(R)  (2005).

\bibitem{SchHofKru05}
T. Schumm {\it et~al.}, Nat. Phys. {\bf 1},  57  (2005).

\bibitem{JoChoPri07}
G.~B. Jo {\it et~al.}, Phys. Rev. Lett. {\bf 98},  180401  (2007).

\bibitem{JonValHin03}
M.~P.~A. Jones {\it et~al.}, Phys. Rev. Lett. {\bf 91},  080401  (2003).

\bibitem{LinTepVul04}
Y.~J. Lin, I. Teper, C. Chin, and V. Vuletic, Phys. Rev. Lett. {\bf 92},
  050404  (2004).

\bibitem{GunKemFor05}
A. G\"{u}nther {\it et~al.}, Phys. Rev. A {\bf 71},  063619  (2005).

\bibitem{WilHofSch05}
S. Wildermuth {\it et~al.}, Nature {\bf 435},  440  (2005).

\bibitem{WilHofBar06}
S. Wildermuth {\it et~al.}, Appl. Phys. Lett. {\bf 88},  264103  (2006).

\bibitem{HalWhiSid07}
B.~V. Hall {\it et~al.}, Phys. Rev. Lett. {\bf 98},  030402  (2007).

\bibitem{HofLesSch07}
S. Hofferberth {\it et~al.}, arXiv {\bf cond-mat},  0710.1575v1  (2007).

\bibitem{AmeEsWic07}
A. van Amerongen {\it et~al.}, arXiv {\bf cond-mat},  0709.1899v1  (2007).

\bibitem{ZobGar01}
O. Zobay and B.~M. Garraway, Phys. Rev. Lett. {\bf 86},  1195  (2001).

\bibitem{ColKnyPerr04}
Y. Colombe {\it et~al.}, Europhys. Lett. {\bf 67},  593  (2004).

\bibitem{WhiGaoDem06}
M. White, H. Gao, M. Pasienski, and B. DeMarco, Phys. Rev. A {\bf 74},  023616
  (2006).

\bibitem{HofLesSch06}
S. Hofferberth {\it et~al.}, Nat. Phys. {\bf 2},  710  (2006).

\bibitem{LesHofSch06}
I. Lesanovsky, S. Hofferberth, J. Schmiedmayer, and P. Schmelcher, Phys. Rev. A
  {\bf 74},  033619  (2006).

\bibitem{KetDru96a}
W. Ketterle and N.~J. van Druten, Advances in Atomic, Molecular and Optical
  Physics {\bf 37},  181  (1996).

\bibitem{JoShiPre07}
G.~B. Jo {\it et~al.}, Phys. Rev. Lett. {\bf 98},  030407  (2007).

\bibitem{ExtLeBSch06}
M.~H.~T. Extavour {\it et~al.},  in {\em Atomic Physics 20}, edited by C. Roos,
  H. H\"affner, and R. Blatt (AIP Conference Proceedings 869, New York, 2006),
  pp.\ 241--249.

\bibitem{HofFisLes07}
S. Hofferberth {\it et~al.}, Phys. Rev. A {\bf 76},  013401  (2007).

\bibitem{HofLesFis07a}
S. Hofferberth {\it et~al.}, Nature {\bf 449},  324  (2007).

\bibitem{GarPerLor06}
C.~L. Garrido~Alzar, H. Perrin, B.~M. Garraway, and V. Lorent, Phys. Rev. A
  {\bf 74},  053413  (2006).

\bibitem{TreGarCor07}
J.-B. Trebbia {\it et~al.}, Phys. Rev. Lett. {\bf 98},  263201  (2007).

\bibitem{BouTreGar07}
I. Bouchoule, J.-B. Trebbia, and C.~L.~G. Alzar, arXiv {\bf cond-mat},
  0707.0602v1  (2007).

\bibitem{LesSchHof06}
I. Lesanovsky {\it et~al.}, Phys. Rev. A {\bf 73},  033619  (2006).

\bibitem{HofLesFis06}
S. Hofferberth {\it et~al.}, Nature Phys. {\bf 2},  710  (2006).

\bibitem{LeaChiKie02}
A.~E. Leanhardt {\it et~al.}, Phys. Rev. Lett. {\bf 89},  040401  (2002).

\bibitem{OttForKra03}
H. Ott {\it et~al.}, Phys. Rev. Lett. {\bf 91},  040402  (2003).

\bibitem{AntPitStr04}
M. Antezza, L.~P. Pitaevskii, and S. Stringari, Phys. Rev. A {\bf 70},  053619
  (2004).

\bibitem{SidDalSch06}
A.~I. Sidorov, B.~J. Dalton, S.~M. Whitlock, and F. Scharnberg, Phys. Rev. A
  {\bf 74},  023612  (2006).

\bibitem{HarObrMcG05}
D.~M. Harber, J.~M. Obrecht, J.~M. McGuirk, and E.~A. Cornell, Phys. Rev. A
  {\bf 72},  033610  (2005).

\bibitem{GueRioGae06}
W. Guerin {\it et~al.}, Phys. Rev. Lett. {\bf 97},  200402  (2006).

\bibitem{WanAndBri05}
Y.-J. Wang {\it et~al.}, Phys. Rev. Lett. {\bf 94},  090405  (2005).

\end{thebibliography}
\end{document}